\documentclass[aps,prd,showpacs,nofootinbib,twocolumn,superscriptaddress,floatfix]{revtex4}

\usepackage{amsmath,amssymb}
\usepackage[dvips, pdftex]{graphicx}
\usepackage[tight,footnotesize]{subfigure}
\usepackage[usenames]{color}
\usepackage[dvipsnames]{xcolor} 
\usepackage{float}
\usepackage{wrapfig}
\usepackage{hyperref}
\usepackage{mathtools}
\usepackage[utf8]{inputenc}
\usepackage{graphicx}
\graphicspath{ {./images/} }

\begin{document}

\newcommand{\tcb}{\textcolor{blue}}
\newcommand{\dd}{\mathrm{d}} 

\def\ga{\mathrel{\raise.3ex\hbox{$>$\kern-.75em\lower1ex\hbox{$\sim$}}}}
\def\la{\mathrel{\raise.3ex\hbox{$<$\kern-.75em\lower1ex\hbox{$\sim$}}}}

\def\be{\begin{equation}}
\def\ee{\end{equation}}
\def\bea{\begin{eqnarray}}
\def\eea{\end{eqnarray}}

\def\betap{\tilde\beta}
\def\del{\delta_{\rm PBH}^{\rm local}}
\def\Msun{M_\odot}
\def\Rcl{R_{\rm clust}}
\def\fPBH{f_{\rm PBH}}

\def\JGB#1{\textcolor{red}{\tt[JGB: #1]}}

\newcommand{\Mpl}{M_P} 
\newcommand{\mpl}{m_\mathrm{pl}} 

\title{A boosted gravitational wave background for primordial black holes \\ with broad mass distributions and thermal features}

\author{Eleni Bagui}

\affiliation{Service de Physique Th\'eorique, Universit\'e Libre de Bruxelles (ULB), Boulevard du Triomphe, CP225, 1050 Brussels, Belgium.}

\author{S\'ebastien Clesse}

\affiliation{Service de Physique Th\'eorique, Universit\'e Libre de Bruxelles (ULB), Boulevard du Triomphe, CP225, 1050 Brussels, Belgium.}

\date{\today}

\begin{abstract}
Primordial black holes (PBHs) with a wide mass distribution imprinted by the thermal history of the Universe, which naturally produces a high peak at the solar mass scale, could explain the gravitational-wave events seen by LIGO/Virgo and up to the totality of the dark matter.  We show that compared to monochromatic or log-normal mass functions, the gravitational wave backgrounds (GWBs) from early PBH binaries and from late binaries in clusters are strongly enhanced at low frequency and could even explain the NANOGrav observations.  This enhancement comes from binaries with very low mass ratios, involving solar-mass and intermediate-mass PBHs at low frequency, solar-mass and subsolar-mass at high frequency. LISA could distinguish the various models, while in the frequency band of ground-based detectors, we find that the GWB from early binaries is just below the current LIGO/Virgo limits and above the astrophysical background, if they also explain black hole mergers.  The GWB from binaries in clusters is less boosted but has a different spectral index than for neutron stars, astrophysical black holes or early PBH binaries. It is detectable with Einstein Telescope or even with the LIGO/Virgo design sensitivity.  

\end{abstract}

\maketitle

\section{Introduction}

Since their first detection in September 2015~\cite{LIGOScientific:2016aoc}, gravitational waves (GWs) have revealed some intriguing properties of black holes:  larger masses than in X-ray binaries, low effective spins, the existence of black holes in the low mass gap or in the pair instability mass gap and binaries with low mass ratios~\cite{LIGOScientific:2018mvr,LIGOScientific:2020ibl}.   In particular, GW190521~\cite{Abbott:2020tfl} involved at least one black hole in the pair instability mass gap.  It could be a secondary merger from a dense cluster~\cite{Abbott:2020mjq} but in such a case, it remains challenging to obtain small enough kick velocities for such a black hole to remain gravitationally bound to the host cluster.  Then, GW190814~\cite{Abbott:2020tfl} combines a primary black hole with an almost zero spin ($\chi_1 <0.07$ at 90\% C.L.) and a probable black hole in the low mass gap.  The merging rates for such event are of the same order than for equal-mass binaries, making it hard to explain with the current stellar models.  In addition, recent microlensing observations by OGLE and Gaia~\cite{Wyrzykowski:2019jyg} towards the galactic center have also revealed a possible population of black holes in the low mass gap.  

Stellar evolution scenarios may be able to explain all those observations, by invoking different black hole populations and binary formation mechanisms.  An alternative attracting a lot of attention is a primordial origin, first suggested in~\cite{Bird:2016dcv,Clesse:2016vqa,Sasaki:2016jop}.   Such primordial black holes (PBHs) may explain a significant fraction or even the totality of the dark matter (DM) in the Universe.  PBHs may have formed in the early Universe with almost vanishing spins due to the gravitational collapse of pre-existing large inhomogeneities~\cite{Hawking:1971ei,Carr:1974nx,1975Natur.253..251C}, for instance produced during inflation.  

Contrary to stellar black holes, PBHs could have any mass above $10^{11}$ kg (otherwise they would have evaporated today through Hawking radiation), but the QCD transition provides a natural explanation for a strong peak in their mass function at the solar-mass scale~\cite{Jedamzik:1996mr,Niemeyer:1997mt,Byrnes:2018clq,Carr:2019kxo}, as well as a bump around $30-100 M_\odot$~\cite{Byrnes:2018clq}.   This could explain the rates, masses and spins of the LIGO/Virgo detections~\cite{Carr:2019kxo,Clesse:2020ghq,Jedamzik:2020ypm,Jedamzik:2020omx}.   These features generically arise from the known thermal history of the Universe and change the equation of state when quarks condensated into protons, neutrons and pions.  

If they compose a significant fraction of the DM, the induced Poisson fluctuations in the PBH density must have seeded PBH clusters at high redshifts~\cite{Kashlinsky:2016sdv,Kashlinsky:2020ial}, resulting in strong modifications of the halo mass function.  This has four interesting effects.  First, PBH clusters induce a suppression of the merging rates of early binaries, below the current limits~\cite{Raidal:2018bbj,Hutsi:2020sol}.  Second, such an effect can explain the observed spatial correlations between the infrared and the source subtracted X-ray backgrounds~\cite{Kashlinsky:2016sdv,Kashlinsky:2018mnu}.  Third, the merging rate of PBH binaries, formed by tidal capture in those clusters, can be at the same level than the rates inferred from GW observations~\cite{Clesse:2020ghq}.  Fourth, most of the PBHs in the halos of massive galaxies should be regrouped in $10^6-10^7 M_\odot$ clusters, which implies that the point sources monitored in microlensing surveys are gravitationally lensed by these clusters.  This reduces the amplitude of microlensing events below the level of detectability, which in turn suppresses the limits on the PBH abundance, allowing $f_{\rm PBH} =1$ at the solar mass scale~\cite{Carr:2019kxo}.  The clustering of PBHs is therefore \textit{unavoidable} (Poisson fluctuations are linked to their discrete nature) and \textit{essential} to evade microlensing and GW limits, if stellar-mass PBHs significantly contribute to the DM.  

If some of the detected black holes were primordial, this would suggest a wide mass distribution, including both subsolar and intermediate mass black holes.  Besides the dark matter, such wide-mass models imprinted by thermal features could explain a series of other puzzling observations~\cite{Clesse:2017bsw,Carr:2019kxo}, including microlensing of stars towards the galactic center due to planetary-mass compact objects~\cite{Niikura:2019kqi}, and of quasars misaligned with the lensing galaxy~\cite{Hawkins:2020zie}.  This could also explain the critical size of ultra-faint dwarf galaxies, the core profile of dwarf galaxies, the existence of supermassive black holes at high redshifts and even the baryon asymmetry of the Universe
~\cite{Garcia-Bellido:2019vlf,Carr:2019hud}.  Studying and searching for other signatures of wide-mass PBH models is therefore important to help distinguish the primordial or stellar origin of black holes. The gravitational-wave background (GWB) from PBH binaries is one of them.

In this paper, we compute for the first time the GWB associated to PBH binaries in wide-mass models including thermal features.  Our analysis therefore extends a previous work that assumed monochromatic and log-normal mass distributions~\cite{Clesse:2016ajp}.  We also consider both early binaries and late-time binaries formed by tidal capture in PBH clusters.  Compared to previous results, we find important differences that are linked to the existence of a significant fraction of light and heavy PBHs in the mass function.  They can combine with PBHs from the QCD peak and form binaries with extreme mass-ratios, contributing to the GWB.   This importantly boosts the amplitude of the GWB at frequencies relevant for LISA and Pulsar Timing Arrays (PTAs).   We find that it can be compatible with the GWB possibly detected by NANOGrav~\cite{2021ApJS..252....4A} and more recently by PPTA~\cite{Goncharov:2021oub}, for binaries in clusters.   A second important finding is the change of the GWB amplitude and spectral index at frequencies relevant for LIGO/Virgo/Kagra (LVK).   The total signal, dominated by early binaries, is found to be within the range of the next observing runs by LVK and above expectations for neutron star and astrophysical black hole binaries.   

Compared to~\cite{Clesse:2016ajp,Mandic:2016lcn,Raidal:2018bbj,Raidal:2017mfl,Mukherjee:2021itf,Mukherjee:2021ags}, our calculations include several improvements: we take into account the rate suppression of early binaries due to Poisson fluctuations and nearby PBHs; we consider merging rates that are consistent with latest GW observations, in particular the ones of GW190425, GW190521 and GW190814.  Most importantly, we consider a wide mass function with thermal effects.  

The paper is organized as follows: in Section~\ref{sec:GWBfromBBH} we briefly introduce the formalism used to obtain the characteristic strain of the GWB from BH mergers in the Newtonian limit. In Section~\ref{sec:massfunction} we present the broad PBH mass spectrum impacted by the thermal history of the Universe. Section~\ref{sec:PBHmerg} focuses on the PBH merging rates of early binaries as well as late-time binaries formed by capture in clusters. In Sections \ref{sec:PBHclust} and \ref{sec:earlyPBH} we compute the GWB for our two binary formation channels and discuss in Sections \ref{sec:GB}, \ref{sec:LISA}, \ref{sec:PTA} and \ref{sec:NANO} its detectability with ground-based interferometers, space-based experiments and PTA's. The last Section is dedicated to outlining and discussing the main results of our analysis and summarizing our conclusions.

\section{GWB from black hole mergers} \label{sec:GWBfromBBH}

A GW background (GWB) is produced by a population of BH binaries experiencing merging and is given by the relative GW energy density $\rho_{{\rm{GW}}}$ to the critical density $\rho_{{\rm{c}}}$ today, per unit interval of logarithmic frequency,
\begin{equation} \label{eq:1}
\Omega_{{\rm{GW}}} \equiv \frac{1}{\rho_{\rm{c}}} \thinspace \frac{{\rm{d}} \rho_{{\rm{GW}}}}{{\rm{d}} \log{f}},
\end{equation}
where $f$ is the observed GW frequency. It is linked to the characteristic GW strain amplitude $h_{\rm{c}}$ through~\cite{Maggiore:1900zz}
\begin{equation} \label{eq:2}
\frac{{\rm{d}} \rho_{{\rm{GW}}}}{{\rm{d}} \log{f}} = \frac{\pi }{4 G} \thinspace f^{2} \thinspace h^{2}_{\rm{c}}(f).
\end{equation}
E.S. Phinney~\cite{Phinney:2001di} established a relationship between the GWB spectrum and the comoving number density of events $N(z)$,
\begin{equation} \label{eq:3}
\frac{{\rm{d}} \rho_{{\rm{GW}}}}{{\rm{d}} \log{f}} = \frac{1}{c^2}\int^{\infty}_{0} N(z) \thinspace \frac{1}{1+z} \thinspace f_{\rm{r}} \frac{{\rm{d}} E_{{\rm{GW}}}}{{\rm{d}} f_{\rm{r}}} \thinspace {\rm{d}}z,
\end{equation}
with $f_{\rm{r}}$ being the frequency of the GWs in a source's cosmic rest frame, related to $f$ by $f_{\rm{r}} = f(1+z)$, $z$ being the redshift and $\dd E_{\rm GW} / \dd f_{\rm r}$ is the redshifted energy produced by each event, per logarithmic frequency interval.
Physically speaking, the GW energy spectrum is a superposition of redshifted GW radiation coming from merging BH binaries over the whole cosmic history.\\
In the Newtonian limit, the energy spectrum associated with two interacting BH binaries is well approximated by~\cite{Maggiore:1900zz}
\begin{equation} \label{eq:4}
\frac{{\rm{d}} E_{{\rm{GW}}}}{{\rm{d}} f_{\rm{r}}} = \frac{\pi^{2/3}}{3G}(G M_{\rm{c}})^{5/3} f_{\rm{r}}^{-1/3}F(e), 
\end{equation}
where $M_{\rm{c}}$ denotes the chirp mass. The function of the orbital eccentricity $F(e)$ is given by~\cite{Clesse:2016ajp},~\cite{Maggiore:1900zz}
\begin{equation} \label{eq:5}
F(e) = \frac{1}{(1-e^{2})^{7/2}} \left(1+\frac{73}{24}e^{2}+\frac{37}{96}e^{4}\right),
\end{equation}
and is close to one in the approximation of circular orbits ($e \simeq 0$).  If one denotes $f_{\rm min}$ the minimum frequency at which this condition is fulfilled, the approximation is valid in the interval $f_{\rm{min}} < f_{\rm{r}} < f_{\rm{{ISCO}}}$, with $f_{\rm{min}}$ being derived from Kepler's law and $f_{\rm{ISCO}}$ corresponding to the innermost stable circular orbit (ISCO), derived from the radius $r_{\rm{ISCO}} = 6 G M/c^{2}$ below which the Newtonian approximation is no longer valid~\cite{Maggiore:1900zz}, that is
\begin{equation} \label{eq:6}
f_{\rm{min}} = \frac{(GM)^{1/2}}{\pi a^{3/2}_{0}}, \hspace{3mm} f_{\rm{max}} \approx 4.4 \thinspace {\rm{kHz}} \left(\frac{M_\odot}{M}\right),
\end{equation}
where $M$ is the total mass of the binary and $a_0$ is the semi-major axis of the orbit.


The number density of GW events $N(z)$ within the redshift interval $[z,z+dz]$ is related to the merging rate $\tau_{\rm{merg}}(z)$ through~\cite{Clesse:2016ajp}
\begin{equation} \label{eq:8}
N(z) = \frac{\tau_{\rm{merg}}(z)}{H(z)(1+z)},
\end{equation}
where $H(z)$ is the Hubble rate, $H(z)= H_{0}\sqrt{\Omega_{\Lambda} + \Omega_{\rm{r}}(1+z)^{4} + (\Omega_{\rm{DM}} + \Omega_{\rm{b}})(1+z)^{3}}$ and $\Omega_{\rm{DM}}$, $\Omega_{\rm{b}}$, $\Omega_{\rm{r}}$ and $\Omega_{\Lambda}$ are  the usual present-day dark matter (DM), baryons (b), radiation (r) and cosmological constant ($\Lambda$) density parameters. 
Throughout the paper, we will consider broad mass distributions that will be introduced in Section~\ref{sec:massfunction}. In this case, the merging rates depend on the two progenitor masses $m_{1}$ and $m_{2}$, so one must integrate over these masses to get the total GWB.
Combining Eqs.~$\eqref{eq:2}$,~$\eqref{eq:3}$,~$\eqref{eq:4}$ and ~$\eqref{eq:8}$ (with $F(e) \simeq 1$), one gets a characteristic strain today
\begin{align} \label{eq:9}
h^{2}_{\rm{c}} (f) &= \frac{4\thinspace  G^{5/3}}{3 \pi^{1/3} \thinspace c^{2}} \thinspace f^{-4/3} \int^{\infty}_{0} \frac{{\rm{d}}z}{(1+z)^{4/3 }H(z)} \nonumber \\
& \times  \int \int {\rm{d}}\ln m_{1} \thinspace {\rm{d}}\ln m_{2} \thinspace \tau_{\rm{merg}}(m_{1}, m_{2}, z) \thinspace M^{5/3}_{\rm{c}},
\end{align}
where the rates must be specified per logarithmic mass intervals.  The PBH mass function with the thermal features enters in these rates and therefore impacts the overall GW background. Justifications for using $F(e) \simeq 1$ in the case of PBH binaries can be found in~\cite{Clesse:2016ajp}.  

\section{Broad PBH mass functions} \label{sec:massfunction}

\begin{figure*}
\includegraphics[scale=0.20]{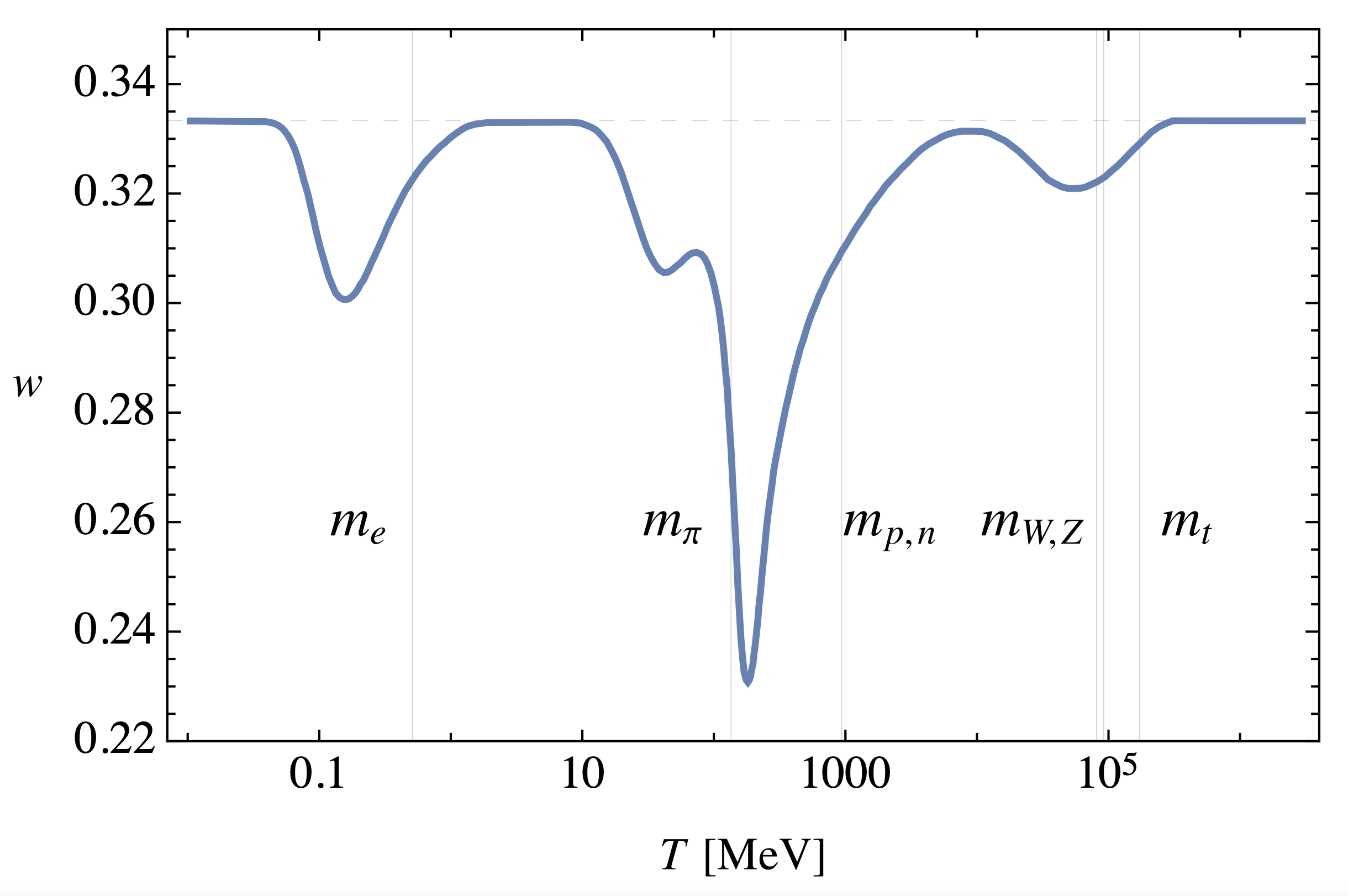} 
\includegraphics[scale=0.094]{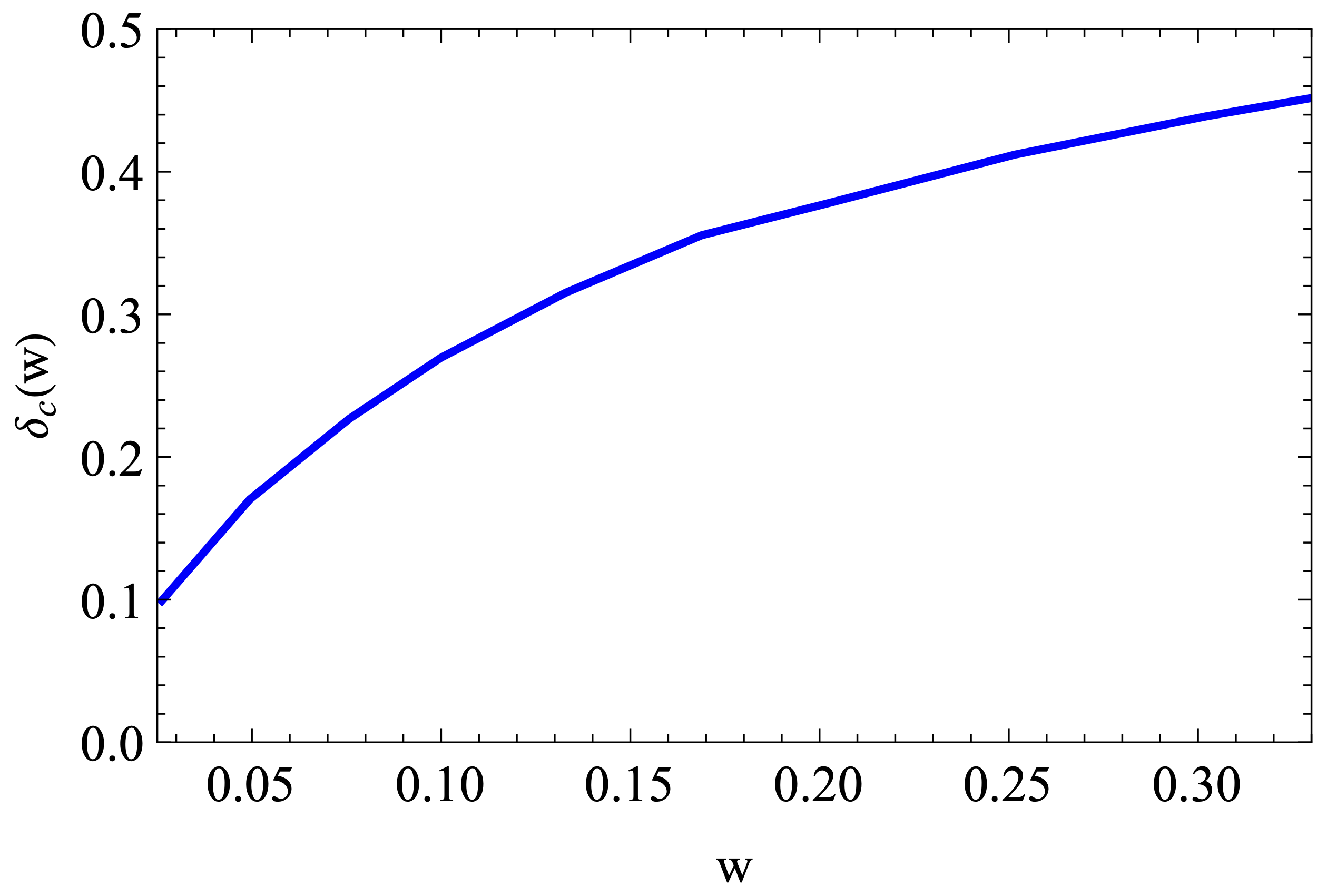}
\\
\includegraphics[scale=0.094]{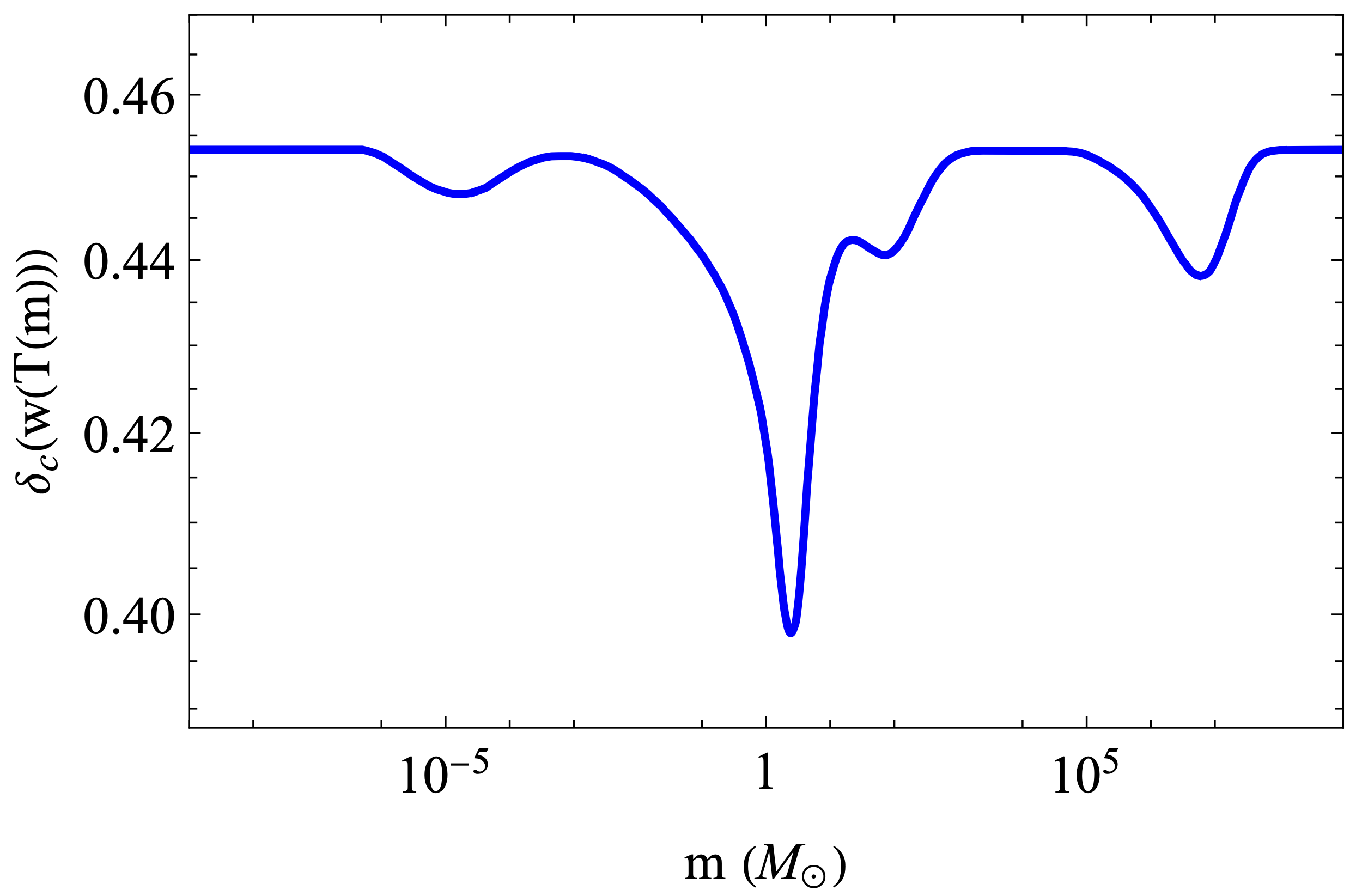}
\includegraphics[scale=0.105]{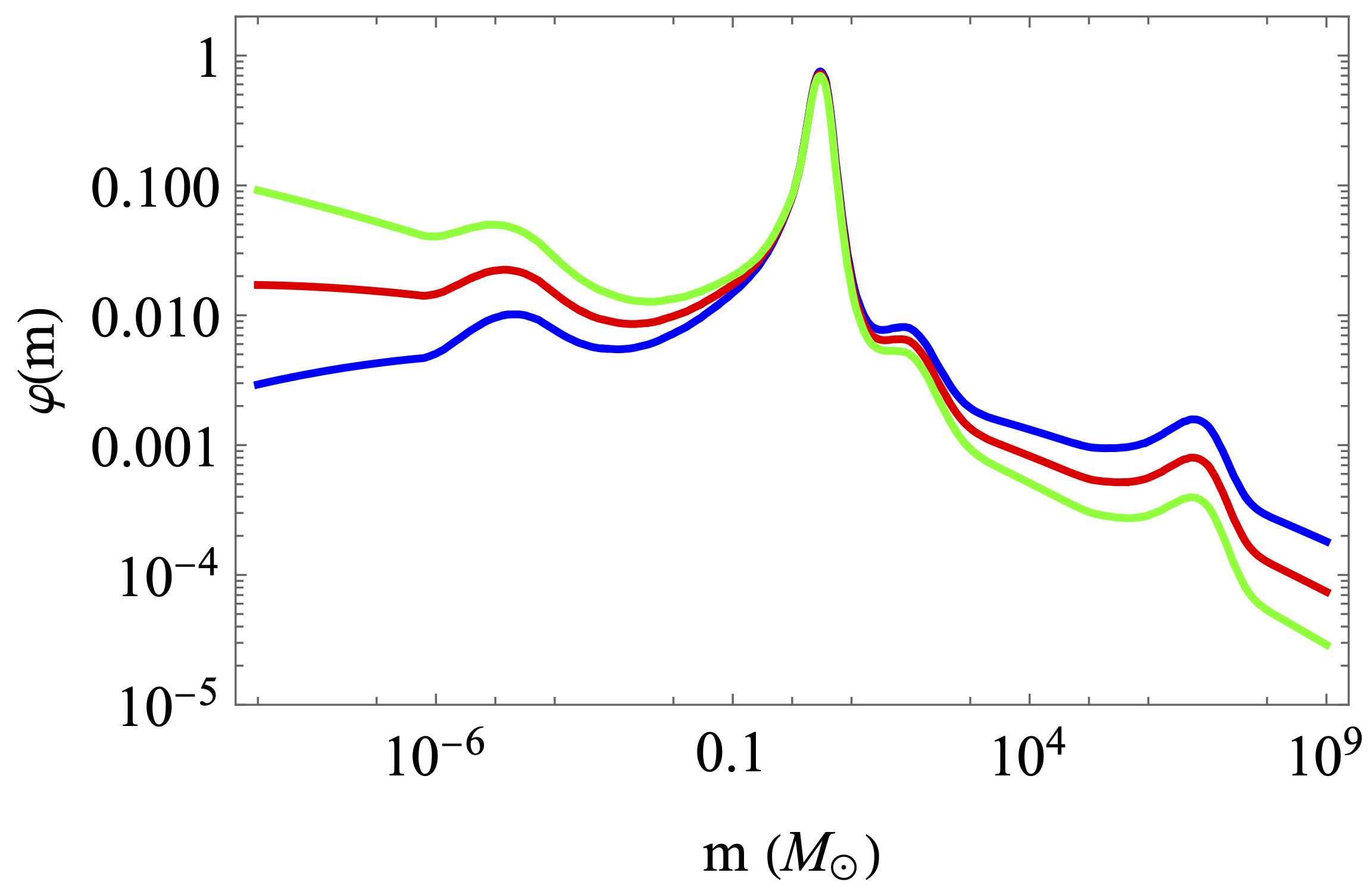}
\caption{Top left: Equation-of-state parameter $w$ as a function of the temperature $T(\rm{MeV})$. The grey vertical lines correspond to the masses of the electron, pion, proton/neutron, W, Z bosons and top quark, respectively. The grey dashed horizontal line indicates the value of $w = 1/3$ (Fig. from~\cite{Carr:2019kxo}).  Top right:  Critical density contrast $\delta_{\rm c}$ as a function of $w$ (data from~\cite{Musco:2012au}). Bottom left: $\delta_{\rm c}$ as a function of the PBH mass (assuming $\gamma = 0.8$). Bottom right:  Resulting broad PBH mass function $\phi(m)$ for a scalar spectral index $n_{\rm s} = 0.965 \thinspace (\rm{dark \thinspace blue}),0.97 \thinspace (\rm{red}),0.975 \thinspace (\rm{green})$ and masses between $10^{-9}$ and $10^{9}M_\odot$.} \label{fig:1}
\end{figure*}

Independently of the DM fraction constituted by PBHs, a primordial origin for some of the observed black hole mergers -- in particular GW190814 that has a low mass ratio -- would imply that PBHs have a somehow extended mass distribution.  Contrary to stellar black holes, their distribution is not restricted by the Chandrasekhar limit on the low-mass side, and by pair-instability on the intermediate-mass side.  Indeed, PBHs could have formed at different times in the cosmic history, possibly including the electroweak, QCD and neutrino decoupling epochs.  The thermal history of the Universe involves a progressive reduction of the relativistic degrees of freedom at these epochs.  This induces slight and transient changes in the equation of state of the Universe and, in turn, in the value of the density threshold for PBH formation.  Unavoidably, this effect induces features in the PBH mass function~\cite{Byrnes:2018clq,Carr:2019kxo}, for the general class of models where PBHs are formed by pre-existing density fluctuations that collapse when re-entering in the horizon during the radiation era (for counter-examples, see e.g.~\cite{Cotner:2018vug, Cotner:2019ykd, Flores:2020drq}). These features take the form of a high peak around the solar mass, coming from the formation of protons and neutrons from quarks, a bump between 20 and 100 $M_\odot$ from the formation of pions, and two other bumps around $10^{-5} M_\odot$ and $10^{6} M_\odot$, respectively coming from the electroweak scale and the neutrino decoupling when electrons and positrons became non-relativistic.

As a result, it becomes possible to obtain a broad and realistic PBH mass function from an almost scale invariant primordial power spectrum, which is a generic outcome of inflation.  One still needs a transition in the power spectrum amplitude between the large cosmological scales and the small PBH scales.  One should note however that in some models, it is possible to form PBHs from rare fluctuations existing on all scales, in a non-Gaussian tail, without changing much the resulting PBH mass function~\cite{Carr:2019hud}, and keeping the primordial power spectrum at the $10^{-9}$ level.  From the point of view of limits on the PBH abundance, using such a mass function is a relatively conservative choice since most models with a peak in the power spectrum could be Taylor expanded on the scales relevant for PBH formation and would then be approximated by an amplitude, a scalar spectral index and a negative running of the scalar spectral index.  We point out that a limit obtained for a null running still applies to a more specific model with a negative running. Therefore, assuming a null running is a conservative approach.

For these reasons, we consider in this paper a small-scale power-law  power spectrum.  
We consider that $\delta_{\rm rms}$, the root-mean-square amplitude of the Gaussian inhomogeneities from which the PBHs have formed, has the following spectrum~\cite{Carr:2019kxo},
\be
\delta_{\rm rms}(m)= A_{\rm s} \left(\frac{m}{M_\odot}\right)^{(1- n_{\rm s})/4}~,
\ee
where $n_{\rm s}$ is a constant small-scale scalar spectral index and $A_{\rm s}$ is the spectrum amplitude whose value is fixed in order to give a total PBH abundance $f_{\rm PBH} \equiv \rho_{\rm PBH} / \rho_{\rm DM}$ today.  $A_{\rm s}$ and $n_{\rm s}$ must not be confused with the usual power spectrum amplitude and scalar spectral index on cosmological scales, inferred from CMB observations.\\
Introducing the fraction $\beta$ of the Universe that collapses to form PBHs of mass $m$ (per unit of logarithmic mass) at the time of formation~\cite{Carr:2019kxo}
\be
\beta(m) = {\rm{erfc}}\left(\frac{\delta_{\rm{c}}}{\sqrt{2}\delta_{\rm rms}(m)}\right),\label{eq:beta}
\ee
and supposing a radiation-dominated Universe until matter-radiation equality, the normalized ($\int \phi(m) {\rm d} \ln m = 1$) broad PBH mass distribution is given by~\cite{Byrnes:2018clq}
\be
\phi(m) \equiv \frac{1}{\rho_{\rm DM} f_{\rm PBH}} \frac{\dd \rho_{\rm PBH}}{\dd \ln m} \approx \frac{2.4}{f_{\rm PBH}} \left(\frac{M_{\rm{eq}}}{m}\right)^{1/2}\beta(m),
\ee
where the numerical factor is $2(1+\Omega_{\rm{b}}/\Omega_{\rm{DM}})$ with $\Omega_{\rm DM} = 0.245$ and $\Omega_{\rm b} = 0.0456$, $M_{\rm{eq}} \simeq 2.8 \times 10^{17} M_\odot$ is the Hubble mass at matter-radiation equality and $f_{\rm PBH}$ is the total dark matter fraction made of PBHs. In Eq.~\ref{eq:beta}, $\delta_{\rm c}$ is the critical overdensity threshold leading to PBH formation, which varies with the equation of state.

Next, we consider the effect of the thermal history and the transient changes in the equation of state parameter $w = p/\rho$ ($p$ being the pressure of the perfect fluid) on $\delta_{\rm c}$.  
Typically, it varies everytime a species (electron, pion, proton, bosons, etc.) decouples from the thermal bath (Fig.~\ref{fig:1}, top left).  Through simulations of PBHs in numerical relativity relating $w$ and $\delta_c$ on the one hand~\cite{Musco:2012au} (Fig.~\ref{fig:1}, top right), and lattice QCD relating $w$ and $T$ on the other hand, one can relate $\delta_{\rm c}$
to the Hubble horizon mass $m_{\rm H}$ (Fig.~\ref{fig:1}, bottom left)~\cite{Byrnes:2018clq}.  The PBH mass is assumed to be $m = \gamma m_{\rm H}$ with $\gamma = 0.8$, as in~\cite{Clesse:2020ghq}, such that the QCD peak lies around $2.5 M_\odot$ where the highest merger rate of compact binary coalescences were observed by LIGO/Virgo. 
The resulting mass functions, for different values of $n_{\rm s}$ (around $0.97$) compatible with LIGO/Virgo observations, are also shown (Fig.~\ref{fig:1}, bottom right).
In the following Sections, we will see that these broad mass functions that take into account the thermal history are included in the expressions of the PBH merging rates, which are used to compute the GWB from PBH mergers.

\section{PBH merging rates} \label{sec:PBHmerg}

PBH binaries can form through two channels: by capture in dense clusters or in the early Universe, when two PBHs form sufficiently close to each other for their gravitational attraction to counterbalance the expansion of the Universe.  This way, they form a binary before the matter-radiation equality. The evolution and mass dependence of the merging rates depend on the formation scenario.   These two channels have been invoked to explain the merging rates inferred by LIGO/Virgo (see e.g.~\cite{Carr:2019kxo,Clesse:2020ghq}).  Nevertheless, both are still subject to large uncertainties, essentially linked to PBH clustering.  Indeed, clustering either induces a suppression or a boost of the merger rates of early binaries and of binaries in clusters, respectively.  Hereafter we summarise some recent prescriptions for these rates, which we assume for the computation of the GWB.  We also briefly discuss the possible connection with observations and the limitations of these approaches.  

\subsection{Early binaries}

An important result of N-body simulations of early PBH binaries is that only a fraction of them are left undisrupted~\cite{Raidal:2018bbj} by other PBHs, clusters or matter fluctuations.  As a result, the merging rates used in~\cite{Sasaki:2016jop} to conclude that the first GW observation impose a limit $f_{\rm PBH} \lesssim 0.01$ at $30 M_\odot$ are in fact overestimated.   Different effects lead to binary disruptions and rate suppression:  i) a close PBH falling into the binary; ii) the binary absorption in an early cluster seeded by Poisson fluctuations; iii)  surrounding large matter fluctuations.  The relative importance of these effects is hard to handle.  Recently, analytical prescriptions have been proposed that are proven to be good approximations when comparing with the results of N-body simulations, for monochromatic or log-normal PBH mass distributions.  Those effects are effectively taken into account by including a suppression factor $f_{\rm sup}$ in the merging rate of early binaries.  As a result, the merger rates of early binaries is given by~\cite{Hutsi:2020sol,Raidal:2018bbj}
\bea
         \frac{\dd\tau_{\rm prim}}{\dd \ln m_1 \dd \ln m_2 } &=& 
        \frac{1.6 \times 10^6}{\rm Gpc^3 yr} \times f_{\rm sup}(m_1,m_2,z) f_{\rm PBH}^{53/37} \nonumber \\   
        & \times &  \phi(m_1) \phi(m_2) \left[\frac{t(z)}{t_0}\right]^{-34/37} \nonumber\\ 
        & \times & \left(\frac{m_1 + m_2}{M_\odot}\right)^{-32/37} \nonumber\\ 
        & \times & \left[\frac{m_1 m_2}{(m_1+m_2)^2}\right]^{-34/37} ~,  \label{eq:cosmomerg}
\eea
where $m_1$ and $m_2$ are the two binary component masses and $t_0$ is the age of the Universe.

The suppression factor $f_{\rm sup}(m_1,m_2,z)$ can be written as the product of $S_1$ and $S_2$, where 
\be
S_1 \approx 1.42 \left[ \frac{(\langle m_{\rm PBH}^2 \rangle/\langle m_{\rm PBH} \rangle^2)}{\bar N + C} + \frac{\sigma_{\rm M}^2}{f_{\rm PBH}^2}\right]^{-21/74} {\rm e}^{-\bar N}
\ee
accounts for binary disruption by matter fluctuations of (rescaled) variance $\sigma^2_{\rm M} \simeq 0.004$ and the number $\bar N$ of nearby black holes that will fall on the binary.  It is given by
\be
\bar N = \frac{m_1+m_2}{\langle m_{\rm PBH} \rangle} \frac{f_{\rm PBH}}{f_{\rm PBH}+\sigma_{\rm M}}.
\ee
Due to the high and relatively sharp QCD peak in the mass function, we assumed that $\langle m_{\rm PBH}^2 \rangle/\langle m_{\rm PBH} \rangle^2 \simeq 1$.  However, we point out that even if subsolar PBHs account for less than a percent of total PBH density, their number density should increase importantly for small masses, which would highly suppress this factor, and thus the merging rates.  It is however hard to estimate the effect of these numerous tiny black holes in the close environment of early binaries made of solar-mass black holes.
Finally, the function $C$ encodes the transition between small and large $\bar N$ limits and can be approximated by 
\bea
C &\simeq& f^2_{\rm PBH} \frac{\langle m^2_{\rm PBH} \rangle / \langle m_{\rm PBH} \rangle^2}{\sigma^2_{\rm M}}  \\
 & \times & \left\{ \left[\frac{\Gamma(29/37)}{\sqrt{\pi}}U\left(\frac{21}{74},\frac{1}{2}, \frac{5 f^2_{\rm PBH}}{6 \sigma^2_{\rm M}}\right)\right]^{-\frac{74}{21}} - 1\right\}^{-1}~. \nonumber
\eea
The second factor $S_2$ comes from the binary disruption in early-forming clusters and can be approximated today by
\be
S_2 \approx \min \left(1,9.6 \times 10^{-3} f_{\rm PBH}^{-0.65} {\rm e}^{0.03 \ln^2 f_{\rm PBH}} \right).
\ee

\begin{figure*}
\includegraphics[scale=0.44]{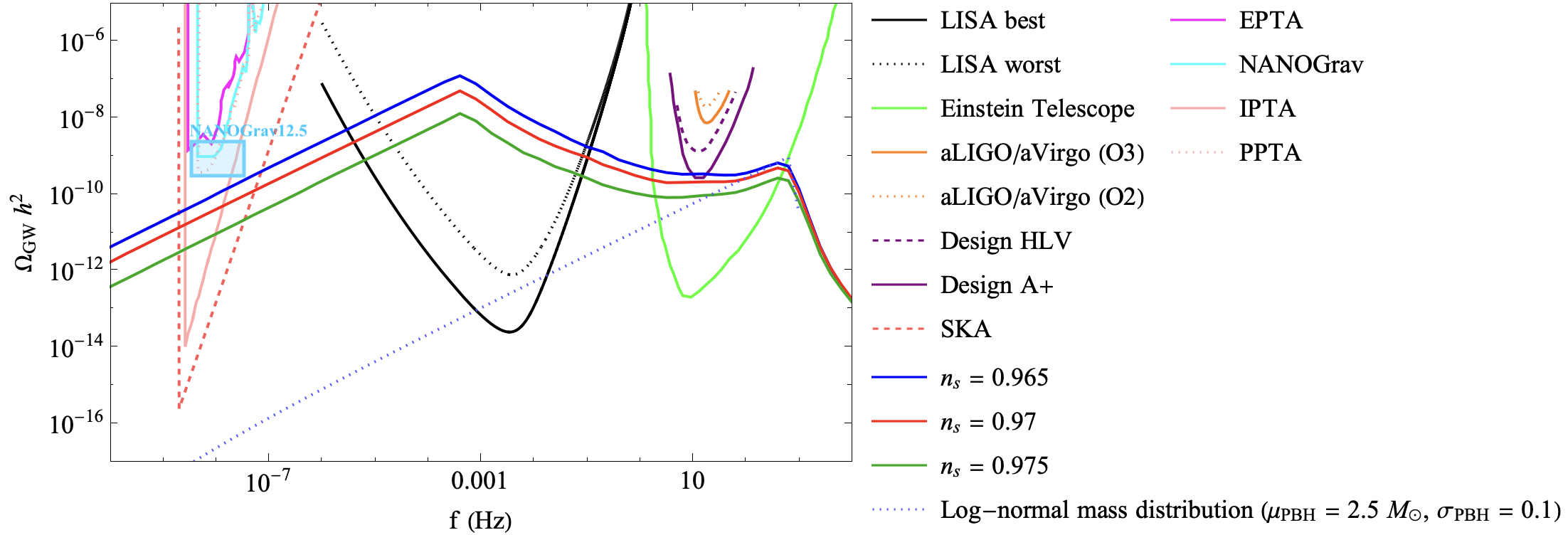}
\centering
\caption{The GW spectrum $\Omega_{\rm GW}h^2$ from late-time PBHs in clusters with a broad mass distribution with scalar spectral index $n_{\rm s} = 0.965 \thinspace {\rm (dark \thinspace blue)}, 0.970 \thinspace {\rm (red)}$ and $0.975 \thinspace {\rm (dark \thinspace green)}$. The late binaries spectrum obtained from a log-normal mass distribution with a mean of $\mu_{\rm PBH} = 2.5 \Msun$ and a standard deviation of $\sigma_{\rm PBH} = 0.1$ is also shown (dotted blue). The numerical spectrum also shows the sensitivities of the ground-based interferometers : LIGO/Virgo O2 (dotted orange) and O3 (solid orange) runs, projections for the HLV (LIGO-Hanford, LIGO-Livingston and Virgo) network at design sensitivity (dashed purple), the mid-scale upgrade of Advanced LIGO (A+ detectors, solid purple)~\cite{PhysRevD.104.022004} and the Einstein Telescope (light green). The sensitivity of the future space-based interferometer LISA is also shown (best experimental design - solid black, worst experimental design - dashed black). The Pulsar Timing Arrays (PTA's) considered here are the EPTA (solid pink), NANOGrav (solid light blue), IPTA (solid light pink), PPTA (dotted light pink), and the Square Kilometer Array (SKA, dashed red)~\cite{SchmitzK}. The recent detection of NANOGrav is represented by the blue square, as in~\cite{DeLuca:2020agl}. In all Figs., we choose a virial velocity of $v_{\rm{vir}} = 5$ km/s of the PBHs in the clusters and density contrasts are such that $R_{\rm clust} = 460$. } 
\label{fig:5}
\end{figure*}

However, in this context, it is important to include the redshift dependence by replacing $f_{\rm PBH}$ by $f_{\rm PBH} \times [t(z)/t_0]^{0.44}$ in the above Equation, valid for $z \leq 100$~\cite{Hutsi:2020sol}. Since the exponential in $S_{2}$ is of order 1 for $0.001 < f_{\rm PBH} \leq 1$, it can be ommited. Moreover, one should  note that if $f_{\rm PBH} \geq 0.0035$, the factor $S_2$ starts to contribute, so one gets
\be
S_2 \approx 10^{-2} \times f_{\rm PBH}^{-0.65} [t(z)/t_0]^{-0.29}.
\ee
That is the approximation we use here. From Eq.~\ref{eq:cosmomerg}, one obtains the merging rates of LIGO/Virgo for masses from the QCD peak, with $f_{\rm PBH}$ close to one.   In particular, for equal-mass mergers and $\bar N = 2$, $f_{\rm PBH} = \phi(m_{\rm PBH}) =1$, one gets $f_{\rm sup} \simeq 0.002$ and merging rates $\tau_{\rm prim} \simeq 5 \times 10^3 (m_{\rm PBH}/M_\odot)^{-32/37} {\rm Gpc}^{-3} {\rm yr}^{-1}$.  This is larger than observed around $30 M_\odot$ but compatible with the rates inferred from GW190425~\cite{Abbott:2020uma} around $2.5 M_\odot$.

Even if the rates given by Eq.~\ref{eq:cosmomerg} are consistent with N-body simulations, there are still a series of uncertainties that may limit this analysis.   First, no N-body simulations have been performed in the case of our broad mass function with thermal features, so it is still possible that (tiny or heavy) PBHs far from the QCD peak additionally suppress these merging rates. Second, Eq.~\ref{eq:cosmomerg} does not take into account the merging rates of the perturbed binaries that may become dominant when $f_{\rm PBH} \gtrsim 0.1$~\cite{Vaskonen:2019jpv}, and so far there are no clear analytical prescriptions for non equal mass binaries.  Third, slightly different results have been obtained in~\cite{Kocsis:2017yty,Liu:2018ess} using analytical methods.
Fourth, it has recently been claimed in~\cite{Boehm:2020jwd} that subtle general relativistic effects may highly suppress this PBH binary formation channel, but this has been disputed in~\cite{DeLuca:2020jug,Hutsi:2021nvs,Hutsi:2021vha,Harada:2021xze}.  Beyond these limitations, one should remember that in general, early binaries can be impacted by their environment (e.g. in clusters or not) during the whole cosmic history.  In particular, our rates for early binaries implicitly assume that only a fraction of PBHs will be in clusters, as opposed to the rate from capture in clusters, typically formed after matter-radiation equality and not considered in N-body simulations of early binaries.  Furthermore, if $f_{\rm PBH} \simeq 1$, evading microlensing limits require that most PBHs are in clusters today.  Therefore, in this case the rate model of early binaries might be somehow incompatible with the late-time formation of clusters.

Strong claims relying on these merging rates are therefore probably still premature. Nevertheless, Eq.~\ref{eq:cosmomerg} probably gives good estimations for some models or regimes, so it is relevant to use it for estimating the GWB, but keeping in mind the possible limitations of this approach.

\subsection{Capture in clusters}

For PBHs in clusters, the merging rate distribution per unit of logarithmic mass can be effectively described by~\cite{Clesse:2020ghq,Clesse:2016vqa,Bird:2016dcv}
\bea 
    \frac{\dd\tau_{\rm clust}}{\dd \ln m_1 \dd \ln m_2 } & = & R_{\rm clust} \times \phi(m_1) \phi(m_2) f_{\rm PBH}^2 \nonumber \\
    & \times & \frac{(m_1 + m_2)^{10/7}}{(m_1 m_2)^{5/7}} \rm{yr^{-1}Gpc^{-3}},
     \label{eq:ratescatpure2}
\eea
where $\Rcl$ is a scaling factor that depends on the PBH clustering properties and velocity distribution, and $\phi(m)$ is the PBH mass function represented in Fig.~\ref{fig:1}. Halo mass functions compatible with the standard $\Lambda$CDM cosmological scenario typically lead to $R_{\rm clust}\approx 1-10 $~\cite{Bird:2016dcv}.  For our mass distribution, this is too low to explain the merging rate inferred from GW190425.  This is also too low to explain the rates around $30 M_\odot$~\cite{Abbott:2016nhf,LIGOScientific:2018jsj}.  However, if $f_{\rm PBH} \simeq 1$ at the solar-mass scale or above, one expects an enhanced clustering due to isocurvature perturbations induced by Poisson fluctuations in the PBH distribution~\cite{Kashlinsky:2016sdv,MoradinezhadDizgah:2019wjf,Clesse:2020ghq}, even though PBHs are not clustered at formation time and follow a Poisson distribution
~\cite{MoradinezhadDizgah:2019wjf}.  This was invoked to explain some spatial correlations in the X-ray and infrared cosmic backgrounds~\cite{Kashlinsky:2016sdv,Kashlinsky:2018mnu,Kashlinsky:2020ial}, and implies that most PBHs reside in dense clusters.  Several physical processes impact the PBH clustering properties:  isocurvature perturbations from Poisson fluctuations, the primordial power spectrum, the dynamical heating and absorption of less massive clusters into larger ones, etc.  These effects naturally lead to a clustering scale around $10^6 M_\odot$ for $m_{\rm PBH} \sim \mathcal O(M_\odot)$, $f_{\rm PBH} \approx 1$ and $R_{\rm clust} \sim \mathcal O(10^2)$, compatible with GW observations.  We assumed that $R_{\rm clust}=460$ as in~\cite{Clesse:2020ghq},  but it is also possible that PBHs are directly formed in clusters, which could induce even larger values of $R_{\rm clust}$.   Finally, let us mention that if most PBHs are regrouped in clusters, the microlensing limits on their abundance can be evaded, allowing $f_{\rm PBH}=1$.

\begin{figure*}[h]
\includegraphics[scale=0.60]{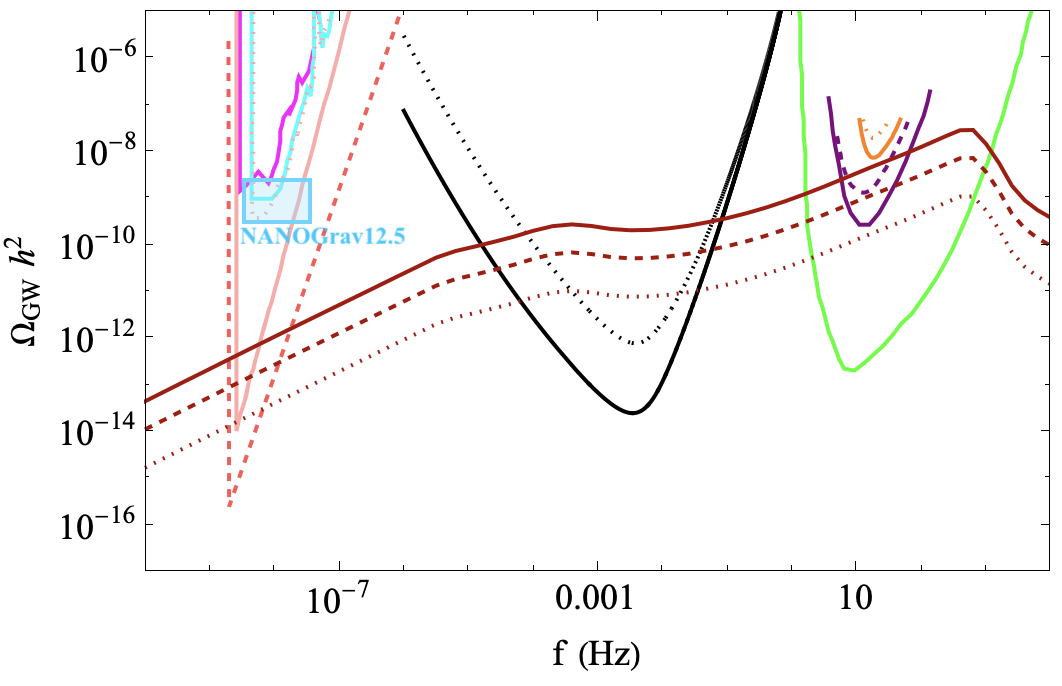}
\includegraphics[scale=0.625]{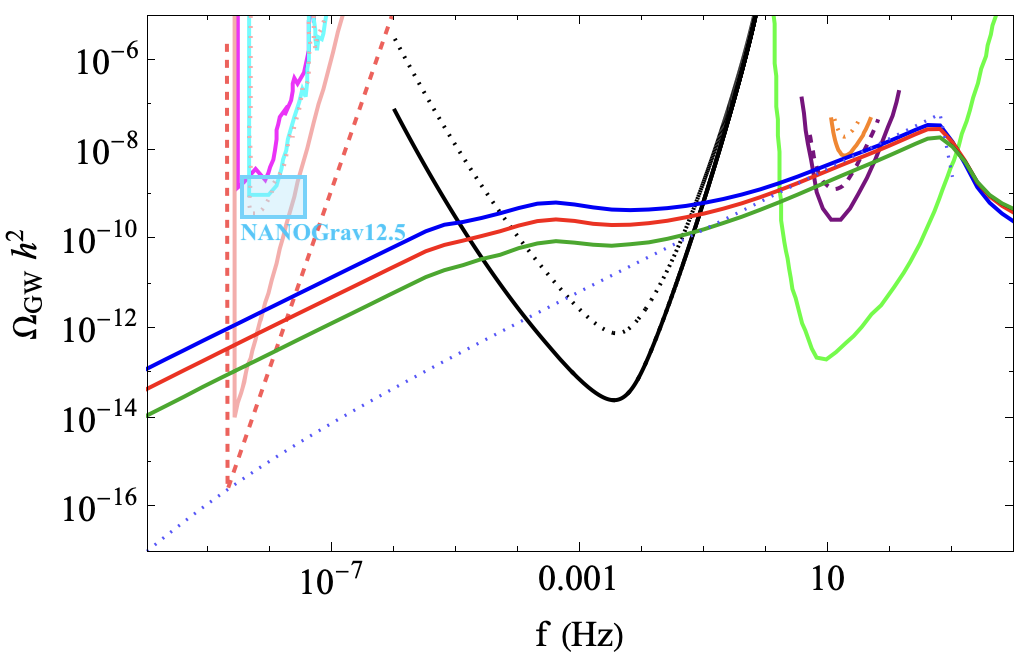}
\centering
\caption{The GW spectrum $\Omega_{\rm GW}h^2$ for early PBH binaries with a thermal broad mass distribution. Upper panel: scalar spectral index $n_{\rm s} = 0.97$ and a PBH abundance $f_{\rm PBH} = 1 \thinspace {\rm ( solid \thinspace dark \thinspace red)}, 0.1\thinspace {\rm(dashed \thinspace dark \thinspace red)}, 0.01 \thinspace {\rm(dotted \thinspace dark \thinspace red)}$. Lower panel : scalar spectral index $n_{\rm s} = 0.965 \thinspace {\rm (dark \thinspace blue)}, 0.970 \thinspace {\rm (red)}$ and $0.975 \thinspace {\rm (dark \thinspace green)}$ and an abundance of $f_{\rm PBH} = 1$. The early binaries spectrum obtained from a log-normal mass distribution with a mean of $\mu_{\rm PBH} = 2.5 \Msun$ and a standard deviation of $\sigma_{\rm PBH} = 0.1$ is also shown (dotted blue). For sensitivity curves, refer to Fig.~\ref{fig:5}.}
\label{fig:6}
\end{figure*}


\section{GWB from PBH clusters} \label{sec:PBHclust}

In this Section, we compute the GW spectrum from late-time PBHs clustered in virialized subhalos of galaxies. To do so, one needs to inject Eq.~$\eqref{eq:ratescatpure2}$ into Eq.~$\eqref{eq:9}$, and the calculation of the characteristic strain is performed by numerical integration on the masses by using a Monte Carlo method, following~\cite{Clesse:2016ajp}.  This allows us to only consider GW frequencies between $f_{\rm{min}}$ and $f_{\rm{max}}$ for our synthetic population of binaries.  Since the integration on the masses is performed numerically, one can write the expression for the characteristic strain without explicitly writing the mass integral,
\begin{align} \label{eq:20}
h^{2}_{\rm{c}}(f) &= \frac{4\tau_{\rm{clust}}}{3 \pi^{1/3}\thinspace c^{2}}\thinspace (G M_{\rm{c}})^{5/3}\thinspace f^{-4/3}  \nonumber \\
&\times \int^{\infty}_{0}\frac{{\rm{d}}z}{(1+z)^{4/3}H(z)}.
\end{align}
The redshift integral is performed separately, since Eq.~$\eqref{eq:20}$ does not contain any other redshift dependence. Expressing the total merging rate in yr$^{-1}$Gpc$^{3}$, the chirp mass in solar masses $M_\odot$ and the frequency in Hz, one gets
\begin{align} \label{eq:21}
h_{\rm{c}} (f) \simeq \thinspace & 1.15 \times 10^{-25} \left(\frac{\tau_{\rm{clust}}}{{\rm{yr}}^{-1} {\rm{Gpc}}^{-3}} \right)^{1/2} \nonumber \\ 
& \times \left(\frac{f}{{\rm{Hz}}}\right)^{-2/3}\left(\frac{M_{\rm{c}}}{M_\odot}\right)^{5/6},
\end{align}
where we used the following result, obtained by numerical integration,
\begin{equation} \label{eq:22}
\int^{\infty}_{0} \frac{{\rm{d}}z}{(1+z)^{4/3 }H(z)} = 0.7642 \thinspace H^{-1}_{0}.
\end{equation}
Considering a synthesized population of $10^{6}$ binaries with given velocities and impact parameters, the marginalized GW spectrum over PBH masses is then obtained. The spectrum is shown in Fig. \ref{fig:5}, where we have chosen the virial velocity in those clusters to be $v_{\rm{vir}} = 5$ km/s (which is compatible with velocity dispersions in faint dwarf galaxies).

\begin{figure*}
\includegraphics[scale=0.49]{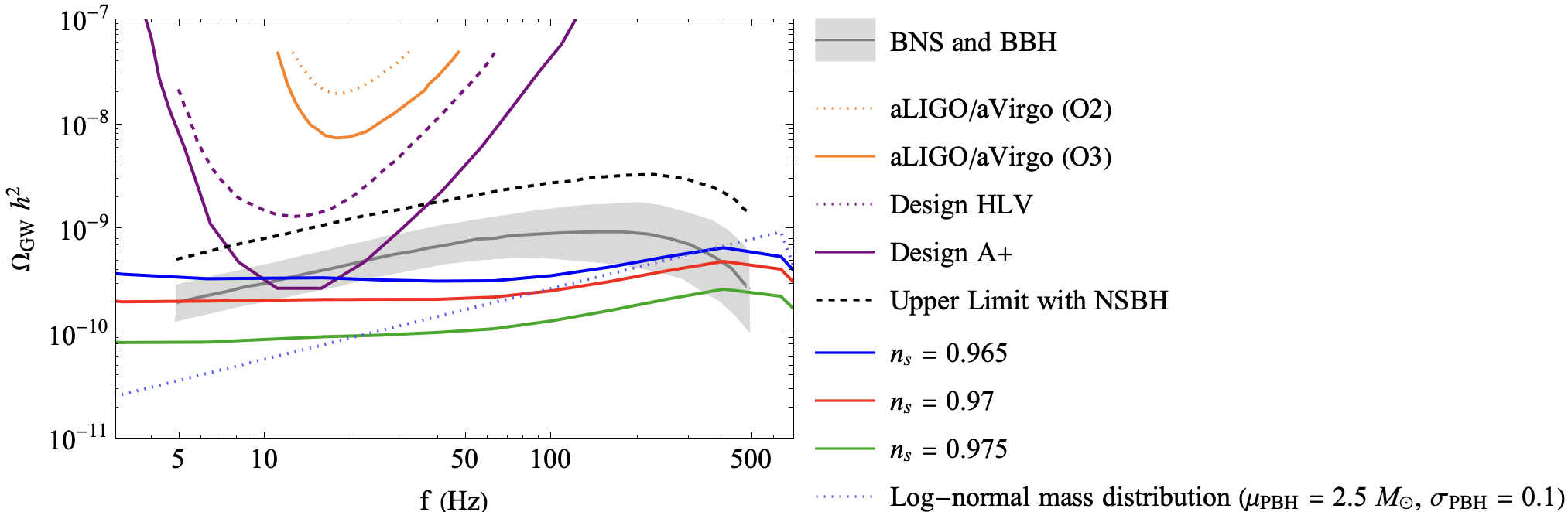}
\centering
\caption{Model predictions for the GWB from BBHs, BNSs and NSBHs, with present and projected sensitivity curves in the frequency range $5-500$ Hz. We have included the combined BNS and BBH energy density spectra, 2$\sigma$ power-law integrated curves for O2 and O3 observing runs (orange lines), projections for the HLV network at design sensitivity and the A+ detectors (purple). The solid gray line shows the median estimate of $\Omega_{\rm BBH+BNS} h^2 (f)$, whereas the shaded gray band represents $90 \%$ credible Poisson uncertainties. The black dashed line indicates the projected upper limit on the total GWB, which includes the upper limit on the contribution from NSBH mergers~\cite{PhysRevD.104.022004}. The GW spectrum $\Omega_{\rm GW}h^2$ from PBH clusters is shown in dark blue ($n_{\rm s} = 0.965$), red ($n_{\rm s} = 0.970$) and dark green ($n_{\rm s} = 0.975$). We choose an abundance of $f_{\rm PBH} = 1$.}
\label{fig:8}
\end{figure*}

\section{GWB from early PBH binaries} \label{sec:earlyPBH}

Early PBH binaries have a merging rate distribution that depends not only on the masses, but also on the redshift $z$ (see Eq.~$\eqref{eq:cosmomerg}$). Consequently, one also has to take into account the redshift term present in the merger rate of early binaries, and perform the redshift integral separately, as in the previous Section. Inserting the primordial rate $\tau_{\rm{prim}}(z)$ in the integral~$\eqref{eq:9}$, one obtains
\begin{align} \label{eq:24}
h^{2}_{\rm{c}}(f) &= \frac{4}{3 \pi^{1/3}\thinspace c^{2}}\thinspace (G M_{\rm{c}})^{5/3}\thinspace f^{-4/3}  \nonumber \\
&\times \int^{\infty}_{0}\frac{{\rm{d}}z}{(1+z)^{4/3}H(z)}\tau_{\rm{prim}}(z),
\end{align}
where $\tau_{\rm{prim}}(z) = \tau_{\rm{prim}} \times (t(z)/t_{0})^{-34/37-0.29}$. One has to perform the following numerical calculation:
\begin{equation} \label{eq:23}
\int^{\infty}_{0} \frac{(t(z)/t_{0})^{-34/37 - 0.29}}{(1+z)^{4/3 }H(z)} \thinspace {\rm{d}}z = 5.92 \thinspace H^{-1}_{0}.
\end{equation}
Using the same units as before, one gets 
\begin{align} \label{eq:24}
h_{\rm{c}} (f) \simeq \thinspace & 3.21 \times 10^{-25} \left(\frac{\tau_{\rm{prim}}}{{\rm{yr}}^{-1} {\rm{Gpc}}^{-3}} \right)^{1/2} \nonumber \\ 
& \times \left(\frac{f}{{\rm{Hz}}}\right)^{-2/3}\left(\frac{M_{\rm{c}}}{M_\odot}\right)^{5/6}.
\end{align}
The GWB produced by these early binaries is shown in Fig. \ref{fig:6}.

\begin{figure*}
 \centering
\includegraphics[width=0.49\linewidth, height=5.8cm]{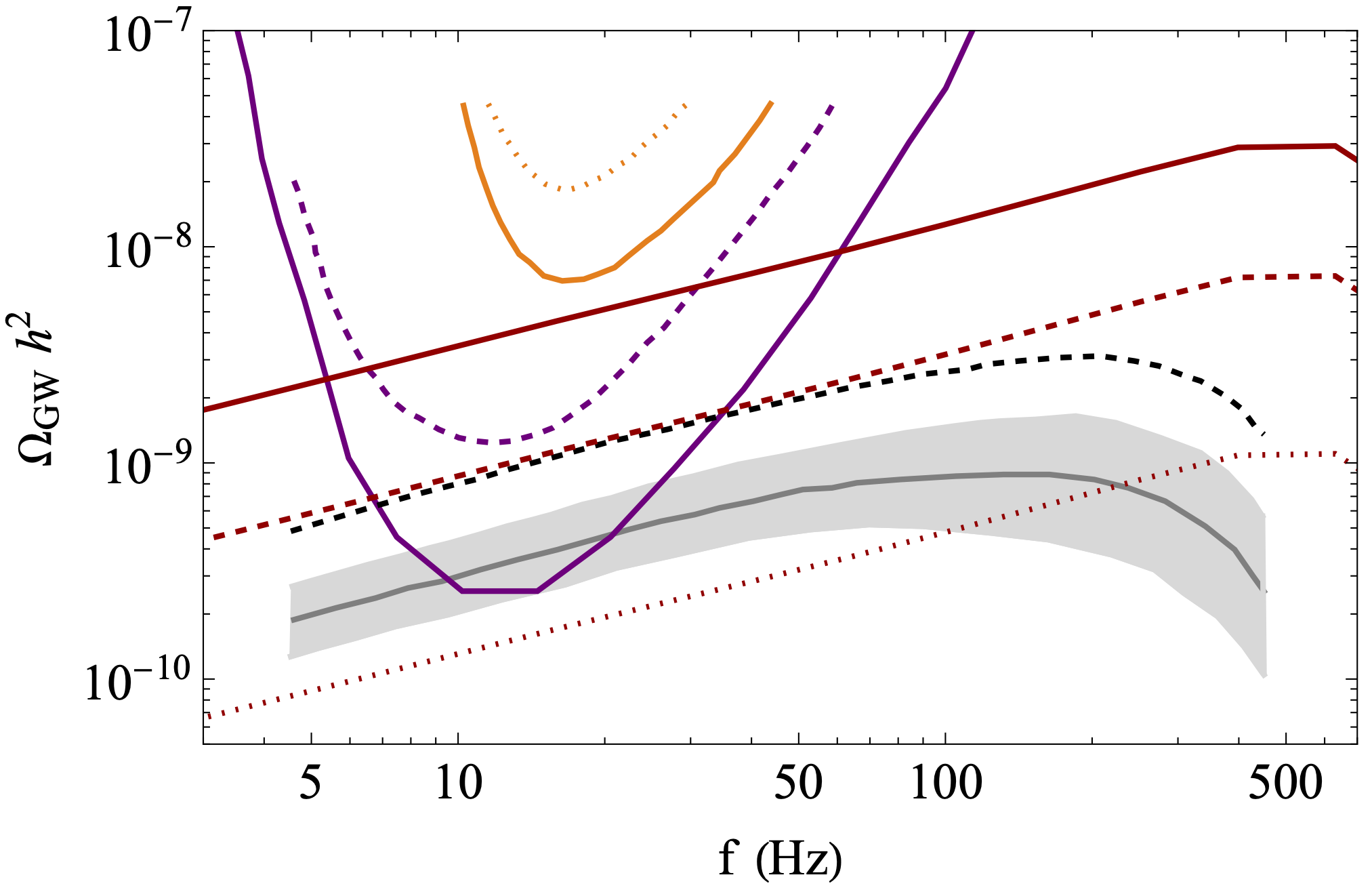} 
\includegraphics[width=0.5\linewidth, height=5.8cm]{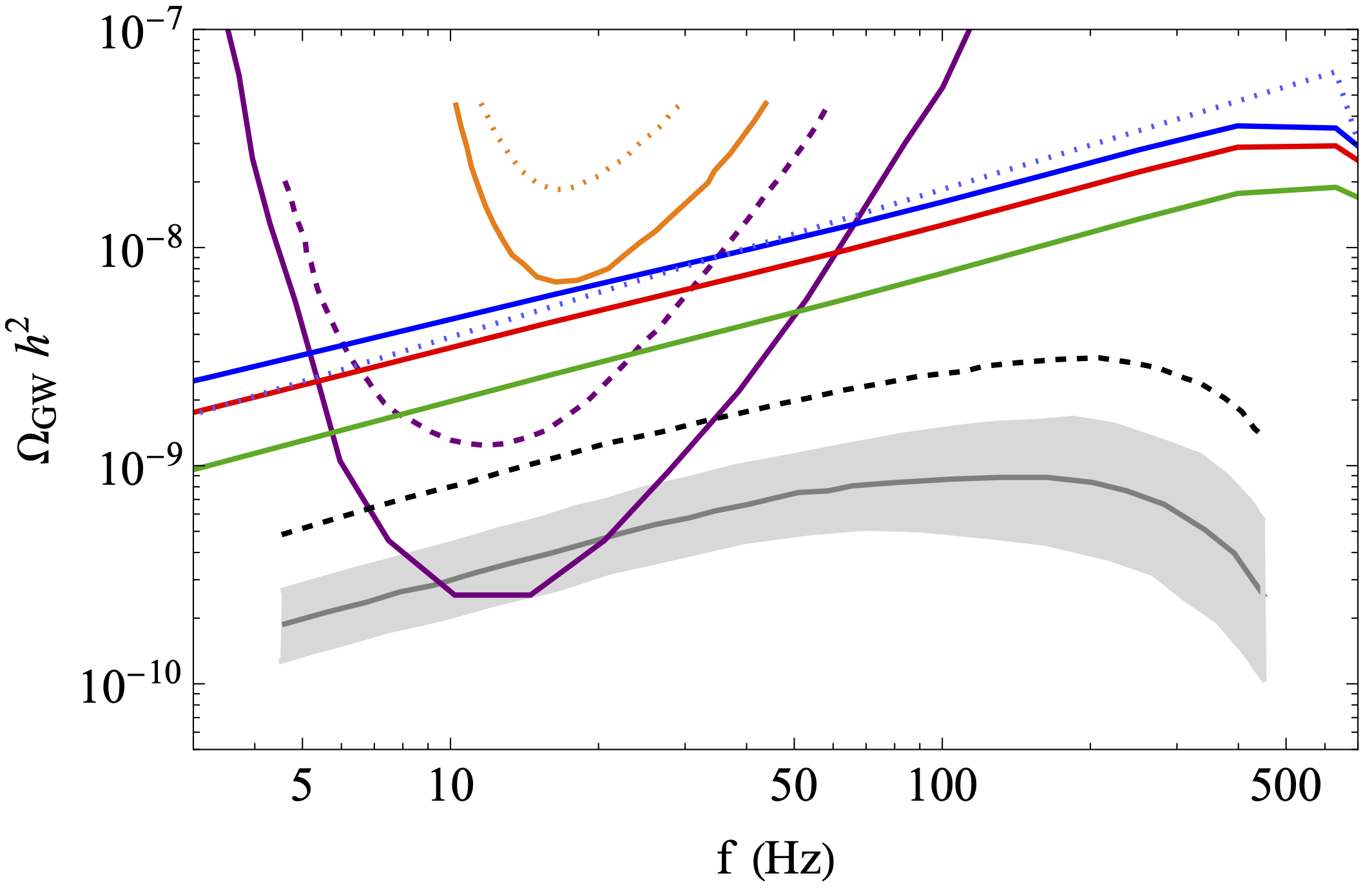}

\caption{Model predictions for the GWB from BBHs, BNSs and NSBHs, with current and projected sensitivity curves in the frequency range $5-500$ Hz (see caption of Fig.~\ref{fig:8}). The GW spectrum $\Omega_{\rm GW}h^2$ from early PBH binaries is shown: (left panel) $f_{\rm PBH} = 1 \thinspace {\rm ( solid \thinspace dark \thinspace red)}, 0.1\thinspace {\rm(dashed \thinspace dark \thinspace red)}, 0.01 \thinspace {\rm(dotted \thinspace dark \thinspace red)}$; (right panel) $n_{\rm s} = 0.965 \thinspace {\rm (dark \thinspace blue)}, 0.970 \thinspace {\rm (red)}$ and $0.975 \thinspace {\rm (dark \thinspace green)}$.}
\label{fig:9}
\end{figure*}

\section{Detectability with ground-based interferometers} \label{sec:GB}

    As shown in Figs.~\ref{fig:5} and~\ref{fig:8}, the $2\sigma$ power-law integrated sensitivity of LIGO/Virgo O2 and O3 observing runs does not yet constrain the expected GW background for PBH clusters.  However, the GW spectrum clearly reaches the sensitivity of Einstein Telescope (ET), as well as the design sensitivity of upgraded LIGO facilities (A+) in the PBH model with the lowest value of $n_{\rm s}$ (0.965).  In Fig.~\ref{fig:8} we also show an updated estimate of the combined GWB coming from the merging of three classes of compact binaries: binary black holes (BBHs), binary neutron stars (BNSs), and neutron star-black hole binaries (NSBHs), from~\cite{KAGRA:2021kbb}. The combined background coming from BBHs and BNSs is predicted to be $\Omega_{\rm BBH+BNS} h^2$(25 Hz)= $3.3^{+3.3}_{-2.3} \times 10^{-10}$. Furthermore, combining the upper limit on $\Omega_{\rm NSBH}(f)$ with the upper 95\% bound on the contributions from BBH and BNS mergers, we set an upper bound on the total expected GWB, $\Omega_{\rm Total} h^{2}$(25 Hz) $\leq 0.86 \times 10^{-9}$.  
    The GWB from PBHs in clusters is approximately at the same level as the one coming from sources of astrophysical origin, though the former is significantly lower around 100 Hz due to merging of PBHs with low mass ratios.  As a result, at the frequency at which LIGO will have its best sensitivity, the GWB spectral index (defined as ${\rm d} \log \Omega_{\rm GW} / {\rm d} \log f$) from PBHs in clusters is slightly negative, whereas the prediction for astrophysical sources is around $2/3$.  Probing this GWB spectral index with ground-based detectors can therefore be used to distinguish an astrophysical from a primordial origin if they have a broad mass function, when one includes thermal features.

    For early PBH binaries, the GWB is represented in Figs.~\ref{fig:6} and~\ref{fig:9} for different values of $n_{\rm s}$ and $f_{\rm PBH}$. We observe that if PBHs contribute to the dark matter with more than $f_{\rm PBH} \approx 0.01$, the GW spectrum is above the projected limits for the HLV (LIGO-Hanford, LIGO-Livingston and Virgo) and A+ detector designs. For $f_{\rm PBH} = 1$, it is just below the current limits imposed by the third observing run of LIGO/Virgo, in particular in the model with the lowest value of $n_{\rm s}$ (0.965). Therefore, the next observing runs (O4 and O5) of LIGO/Virgo may be crucial for its detection and for the determination of its properties. The spectrum is also well above expectations for astrophysical sources if the PBH constitute almost all of the dark matter, whereas for lower values of $f_{\rm PBH}$, the GW spectrum crosses the upper limit on the GWB from BBHs, BNSs and NSBH mergers, and has the same spectral index. In this case, differentiating the two backgrounds in future detections could be challenging. 
    
    
 
 Our analysis also reveals the important contribution to the GWB of PBH binaries with very low mass ratios, combining a solar-mass PBH from the QCD peak with a subsolar mass PBH.   In order to quantify the contribution form the different masses, we have computed the GWB amplitude $\Omega_{\rm GW} h^2$ as a function of $m_1$ and $m_2$, both for clusters and early binaries, shown in Fig.~\ref{fig:10}.  This way, we explain why the dark green curve corresponding to $n_{\rm s} = 0.975$ (Figs.~\ref{fig:5} and~\ref{fig:6}) is lower than for the other values of $n_{\rm s}$, given that massive black holes are less numerous (see Fig.~\ref{fig:1}) and so contribute less to the GWB.  Moreover, as shown in Fig.~\ref{fig:6}, increasing the PBH abundance $f_{\rm PBH}$ increases the amplitude of the GW spectrum. This can be seen directly from Eq.~$\eqref{eq:cosmomerg}$, and comes from the fact that a higher PBH abundance makes a greater contribution to the spectrum.   At frequencies above 100 Hz, binaries with subsolar black holes dominate the signal, leading to an opposite conclusion.  This particular behaviour could be probed by ET. Finally, we point out that the relative contribution of binaries with low-mass ratios and a subsolar black hole is more important for early binaries than for PBHs in clusters.

\section{Detectability with LISA} \label{sec:LISA}

\begin{figure*}[h]
\includegraphics[scale=0.52]{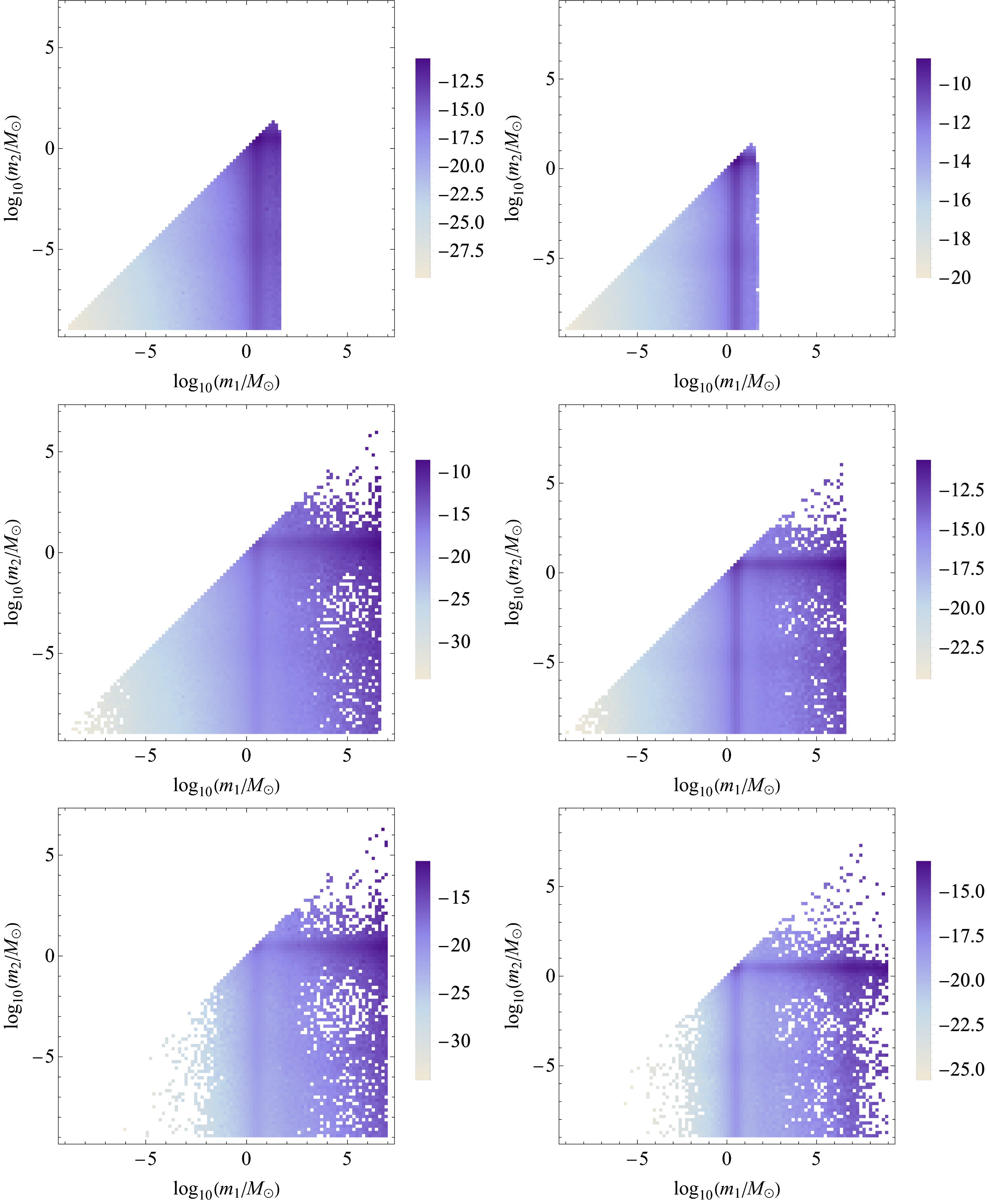}
\centering
\caption{The mass contribution to the GWB from late-time and early PBHs binaries. The color bar indicates the values of the quantity $\log_{10}(\Omega_{\rm GW}h^{2})$, which is represented as a function of the logarithm of the masses $m_1$ and $m_2$ of the PBHs following a mass distribution from the thermal history of the Universe (with $n_{\rm s}$ = 0.97 and $v_{\rm vir}$ = 5 km/s) for the frequencies $f = 100$ Hz (top), $10^{-3}$ Hz (middle) and $10^{-8}$ Hz (bottom). Left column: late-time PBHs binaries in clusters. Right column: early PBHs binaries (with $f_{\rm PBH} = 1$).}
\label{fig:10}
\end{figure*}

Contrary to the case of a log-normal, close to monochromatic mass distribution ($\sigma_{\rm PBH} = 0.1$), the case of a broad PBH mass distribution with thermal features exhibits a GWB with a negative spectral index (${\rm d} \log \Omega_{\rm GW} / {\rm d} \log f < 0$) down to a frequency around $5\times10^{-4}$ Hz, leading to a strong enhancement (by up to 6 orders of magnitude compared to the log-normal mass function) in the frequency range of LISA, for late-time PBHs in clusters (Fig.~\ref{fig:5}). It would be detectable in a broad range of frequencies ($10^{-6}$ to $0.1$ Hz) even for the worst experimental design, allowing for a precise reconstruction. This strong increase in amplitude comes from the merging of binaries with large mass ratios involving an intermediate-mass and a solar-mass (QCD) PBH, as shown in Fig.~\ref{fig:10}. This will help differentiate this background from the one of astrophysical origin.  A similar feature is observed in the GWB for early binaries, shown in the lower panel of Fig.~\ref{fig:6}, except that the amplitude of the background is up to two orders of magnitude lower than for PBHs in clusters, and that the GWB spectral index remains overall positive.  Hence, probing the GWB spectral index at these frequencies with LISA could in principle help us differentiate late-time PBH binaries in clusters from early PBH binaries. Finally, one can notice a bump around $10^{-3}$ Hz, coming from the bump in the PBH mass function around $10^{6} M_\odot$.  The shape of the GWB could therefore be used to reconstruct the exact PBH mass function at large mass.

\section{Detectability with PTA's} \label{sec:PTA}

Another difference coming from the above-mentioned effect is that the GWB induced by PBHs in clusters with a broad mass distribution and consistent with LIGO/Virgo merging rates is now detectable by the PTA's (Fig.~\ref{fig:5}), in particular by future SKA and IPTA observations, as opposed to the log-normal case where the amplitude is very strongly suppressed at low frequency. 
We also find that the change of the GWB spectrum with $n_{\rm s}$ is more important than at higher frequencies. This is due to the fact that the QCD peak remains unchanged as we vary $n_{\rm s}$ (see Fig.~\ref{fig:1}) while the abundance of intermediate-mass PBHs strongly changes. The latter dominate the GWB signal at nanoHertz and therefore the three curves are shifted away from each other. 

\section{A hint or a limit from NANOGrav?} \label{sec:NANO}

The NANOGrav collaboration recently published an analysis of 12.5 years of pulsar timing (PT) data which provides evidence of a stochastic process that can be interpreted as a background of GWs~\cite{2021ApJS..252....4A}.  A similar observation has been recently reported using data from Parkes PTA's (PPTA)~\cite{Goncharov:2021oub} and a possible non-GW origin was also explored.   The detected signal is within the interval $10^{-9} - 10^{-8}$ Hz with $\Omega_{\rm GW}h^{2} \sim 10^{-9}$ (blue window).  The density fluctuations at the origin of solar-mass or planetary-mass PBHs have already been considered as a source of such a GWB~\cite{DeLuca:2020agl,Chen:2019xse,Kohri:2020qqd,Domenech:2020ers}. 


Our analysis reveals that the signal, when interpreted as a GWB, could also be explained by binaries of PBHs with an extended mass function. While the NANOGrav signal does not exactly coincide with the curves shown in Fig.~\ref{fig:5} for PBH binaries in clusters, this would have been the case for a scalar spectral index smaller than 0.965.
As for the early PBH binaries, the amplitude of the GWB is always much lower than the NANOGrav signal. Consequently, on the basis of our results, the observed signal could be explained by PBH binaries in clusters with very small mass ratios (with a primary black hole of intermediate mass) and a small enough spectral index ($n_{\rm s} < 0.965$), if PBHs constitute the totality or a large fraction of the DM.

If the GW nature of the NANOGrav signal were to be confirmed, the measurement of its spectral index~\footnote{Encoded in the so-called $\gamma=5-n=3-2\alpha$ parameter in~\cite{2021ApJS..252....4A} where $n$ is the GWB spectral index and $\alpha$ is the strain spectral index.} could distinguish between a background from PBH binaries or from second order perturbations (almost flat in our model)~\cite{Clesse:2018ogk}.


\section{Conclusion} \label{sec:conclusion}

 We have studied the GWB induced by the formation and merging of binaries from a PBH population. In this work, we only considered PBH-PBH mergers, but one could also explore the GW background from PBH-neutron star mergers.  Nevertheless, their rates are expected to be subdominant for the models we considered here~\cite{Sasaki:2021iuc}.  For the first time, we took into consideration wide PBH mass distributions with features from the thermal history.
 We calculated and characterized the GWB produced by a superposition of GW sources coming from the merging of PBHs binaries in clusters and early PBH binaries, and compared it to the astrophysical GWB produced by astrophysical black holes, binary neutron stars and neutron star-black hole binaries. We then confronted it to the latest limits of LIGO/Virgo and PTA's, and to the forecasted limits of the Einstein Telescope and LISA.  

For binaries in clusters, we found that the GWB amplitude is greatly enhanced than in previous estimations for monochromatic or log-normal mass functions at frequencies below $10$ Hz.  We identified the origin of this boosted GWB, coming from the numerous binaries with low mass ratios. This GWB will be probed by future ground and space-based facilities such as A+, ET or LISA.  We expect that this feature will allow us to distinguish this background from the astrophysical GWB because of the specific frequency dependence.  Interestingly, the GWB at nanoHertz frequency could coincide with a signal from NANOGrav for a small enough spectral index.

 
 For early binaries, we used the most recent merger rate prescriptions, and found that it is also possible to have a wide mass function that can explain GW events and, at the same time, pass the most recent LIGO/Virgo limits on the GWB. This background typically dominates by one order of magnitude the ones from PBH clusters and astrophysical binaries. Nevertheless, the complexity of the phenomena still poses important challenges and one probably still needs to better characterize the merging rates for the early binaries by taking into consideration broad mass functions with thermal features, non-equal mass binaries, gas accretion, relativistic effects, etc., and verify their consistency with N-body simulations.
 
 
 

To summarize, our work has revealed strong differences between PBH models and the possiblity to discriminate the different mass functions and binary origins with the observations of the GWB.  In this context, multi-wavelength observations are essential.   Future instruments like Einstein Telescope, LISA and SKA will play a major role and will be able to confirm or exclude a primordial origin of black holes.

\acknowledgements{The authors warmly thank Juan Garc\`ia-Bellido, Matteo Braglia, Sachiko Kuroyanagi, Mairi Sakellariadou, Alex Jenkins and Federico De Lillo for useful comments and discussion.  E.B. is supported by the FNRS-IISN (under grant number 4.4501.19 ).  S.C. acknowledges support from the Francqui Foundation.  }

\bibliographystyle{apsrev4-1}

\bibliography{PBH_biblio} 

\end{document}